\documentclass[aps,prl,twocolumn,groupedaddress]{revtex4-2}  
\usepackage{graphicx}
\usepackage{amssymb}
\usepackage{amsmath}
\usepackage{newtxtext,newtxmath,lipsum}
\usepackage{xcolor}

\usepackage[shortlabels]{enumitem}
\DeclareFontFamily{U}{matha}{\hyphenchar\font45}
\DeclareFontShape{U}{matha}{m}{n}{
	<5> <6> <7> <8> <9> <10> gen * matha
	<10.95> matha10 <12> <14.4> <17.28> <20.74> <24.88> matha12
}{}
\DeclareSymbolFont{matha}{U}{matha}{m}{n}

\DeclareMathSymbol{\Lt}{3}{matha}{"CE}
\DeclareMathSymbol{\Gt}{3}{matha}{"CF}

\begin{document}
\bibliographystyle{revtex}

	\title{Normal mode splitting induced synchronization blockade in coupled quantum van der Pol oscillators}

	\author{Nissi Thomas}
	\affiliation{Department of Nonlinear Dynamics, School of Physics, Bharathidasan University \\ 
		Tiruchirappalli - 620024, Tamil Nadu, India.}
	
	\author{M. Senthilvelan}
	\affiliation{Department of Nonlinear Dynamics, School of Physics, Bharathidasan University \\ 
		Tiruchirappalli - 620024, Tamil Nadu, India.}

	\begin{abstract}
		
		We report a normal-mode induced  synchronization blockade in coupled quantum van der Pol oscillators under the influence of external drive. In this mechanism, the coupling hybridizes the oscillator modes into spectrally split normal modes. The destructive interference between the transitions to these modes blocks synchronization. We find that this blockade can be controlled simply by tuning the coupling strength and detuning allowing dynamic manipulation of quantum synchronization through collective mode dynamics. We analyze the phase-locking behaviour using perturbation analysis. Further, by deriving steady-state probability amplitudes we show how the energy redistribution and spectral splitting forms the basis of the blockade. Our results might provide new insights into how synchronization can be controlled in quantum systems.
			
	\end{abstract}

\maketitle
Synchronization is an emergent effect which occurs in a group of self-sustained oscillators with random frequencies often mediated by coupling, noise or external forcing. It is a diverse phenomenon evident from natural to engineered systems \cite{pikovsky}. It has been thoroughly understood classically and recently it has emerged as a significant area of research in quantum domain \cite{goychuk, zhirov, giorgi, lee, walter14, walter, lee2, xu, ameri, amitai, weiss, hush, nissi, li1, li2, chia, kato, murtadho}. Synchronization mechanism varies qualitatively in quantum regime from its classical implementation \cite{weiss2, lorch, mari, roulet, sonar} and has been investigated in different quantum systems such as optomechanical systems \cite{amitai, weiss, fan, liao, qiao}, nanomechanical systems \cite{matheny}, super-conducting circuits \cite{hriscu}, micromasers \cite{rodrigues, tilley}, spin system \cite{roulet, roulet2, laskar, tindall}, atomic ensembles \cite{xu, zhu, cabot} and so on.

Quantum synchronization blockade is a nonclassical effect in quantum systems that inhibits or suppresses synchronization. This is in contrast to classical systems where coupling generally promotes synchronization. Blockade can arise from frequency shifts in the energy spectrum \cite{lorch}, interference effects linked to dissipation \cite{roulet2, koppenhofer} and symmetry constraints \cite{solanki, tan, kehrer1}.  More recently, the blockade mechanism emerging from non-reciprocal coupling was analysed in Ref. \cite{kehrer2}. Differing from the earlier works, in this work we report a new normal-mode-induced synchronization blockade mechanism where coupling hybridizes the oscillator modes into spectrally split normal modes in coupled quantum van der Pol oscillators. Unlike the blockade mechanisms reported earlier in the literature,  normal mode splitting-induced blockade results from collective mode hybridization and destructive interference between these split modes.  This mechanism is fundamentally distinct as it allows the  synchronization blockade to be controlled by tuning the system parameters like coupling strength and detuning frequency thereby demonstrating that blockade can be achieved and manipulated through the collective quantum dynamics of the system.

Normal mode splitting emerges in coupled systems as a result of constructive and destructive interference between the modes \cite{junker}. This is a key feature of a strong coupling, where the energy between the modes surpasses the dissipation leading to collective hybrid modes \cite{gblacher}. This regime is crucial to understand the coherent quantum dynamics between the interacting systems with the objective of enabling precise control over the quantum systems \cite{ockeloen, safavi}.  This effect has also been demonstrated in various experimental platforms such as optomechanical \cite{boca, shahidani, wu, rossi},  atom-cavity \cite{thomson} and quantum dot microcavity systems \cite{rjp} when they are strongly driven or kept at cryogenic conditions allowing quantum control. 

In this work, we explore the synchronization mechanism, specifically when the interaction term facilitates the exchange of linear excitation in coupled quantum van der Pol oscillators under the influence of external drive. We show that normal mode splitting, arising from the constructive and destructive interference between the oscillator modes, induces a synchronization blockade. Quantum interference mechanisms play a central role in several pioneering works, including photon blockade driven by interference in photonic cavities \cite{bamba}, dark state formation enabling weak-light nonlinearity in quantum dots through engineered dissipation \cite{yokoshi}, collective synchronization emerging from mode hybridization in optomechanical arrays \cite{ludwig}, and fundamental cooling limits governed by destructive interference in optomechanical systems \cite{weiss3}. Thus, in these works, interference predominantly leads to blockade or suppression of transitions in platforms such as photonic cavities and quantum dots \cite{bamba, yokoshi}, and sets limits on cooling in optomechanical systems \cite{weiss3}, whereas constructive interference can actively promote synchronization by aligning the phases in optomechanical arrays \cite{ludwig}. Our study focuses on the onset and suppression of mutual phase synchronization induced by this interference mechanism. Specifically, conventional synchronization is observed in the low-coupling regime, whereas synchronization blockade occurs when the coupling surpasses the linear dissipation rate, driven by interference-induced normal mode splitting. This reveals that in our system, quantum interference-manifested through normal mode hybridization governs not only excitation transfer, but can also suppress collective phase-locking between oscillators by inducing synchronization blockade. To understand how the drive and coupling affects the phase-locking mechanism, we analyze the phase-locking behaviour in the considered system through perturbation method. To investigate the energy redistribution between the oscillator modes, we analytically obtain the steady-state probability amplitudes and demonstrate how the steady states uncovers the spectral splitting arising from the interference mechanism. Based on these results we illustrate that the synchronization can be actively tuned through coupling strength, thereby enabling quantum control.   

We consider reactively coupled quantum van der Pol oscillator where the first oscillator is subjected to the external drive. The Lindblad master equation describing such a system takes the following form \cite{lee}
\begin{equation}
\dot\rho=-i[H_0+H_I,\rho]+\sum_{i=1}^{2}\gamma_1\mathcal{D}[a_i^{\dagger}]\rho+\gamma_2\mathcal{D}[a_i^2], \label{mas_coup}
\end{equation}
with Hamiltonian, $H_0=\sum_{i=1}^{2}\Delta_ia_i^{\dagger}a_i$, where $a_i$ ($a_i^{\dagger}$) are the bosonic annihilation (creation) operators of the $i^{th}$ oscillator and the Lindblad operator $\mathcal{D}[\hat{a}]\rho=\hat{a}\rho\hat{a}^{\dagger}-\frac{1}{2}\{\hat{a}^{\dagger}\hat{a}\rho+\rho\hat{a}^{\dagger}\hat{a}\}$ describes the non-unitary dynamics of the system. The system undergoes two incoherent processes: negative linear damping with rate $\gamma_1$ and nonlinear damping with rate $\gamma_2$. The Hamiltonian, $H_I$ is comprised of a coherent coupling  and a drive term acting on the first oscillator which can be described in the following form in the rotating frame of the drive \cite{lee},
\begin{eqnarray} 
H_I&=&E(a_1+a_1^{\dagger}) + V(a_1^{\dagger}a_2+a_1a_2^{\dagger}),\label{ham}
\end{eqnarray}
where $E$ is the drive strength, $V$ is the coupling strength and $\Delta_i=\omega_i-\omega_e$ denotes the frequency detuning between the $i^{th}$ oscillator ($\omega_i$) and the drive  ($\omega_e$).
\begin{figure}
	\centering
	\includegraphics[width=0.5\textwidth]{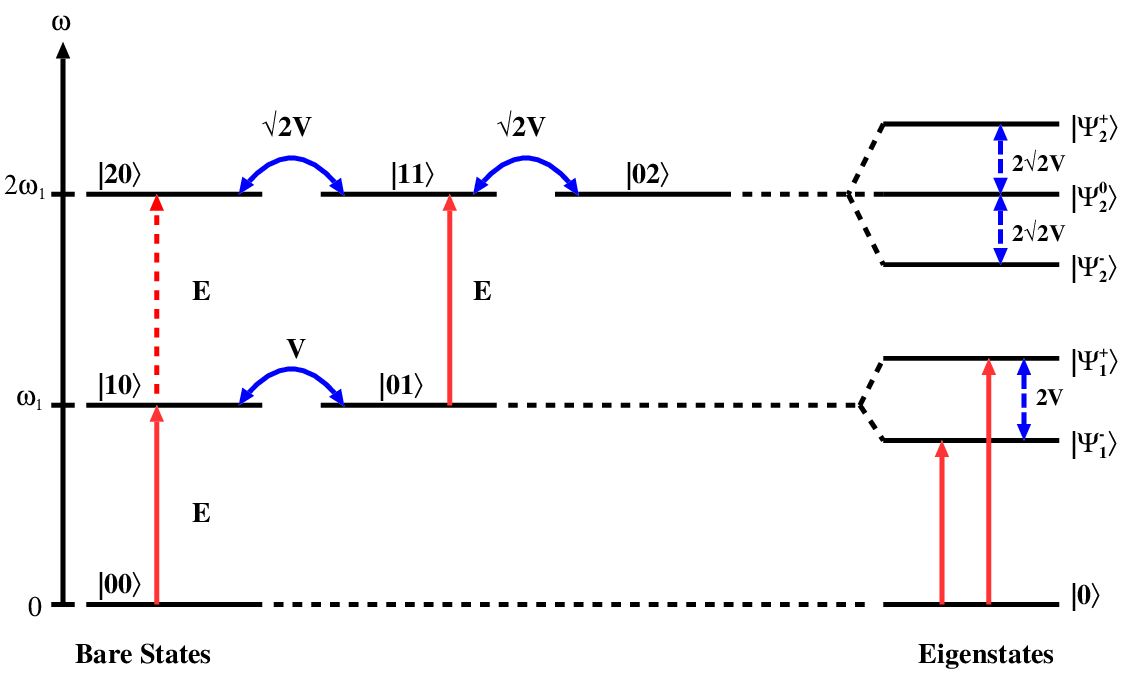}
	\caption{Schematic energy level diagram of the coupled quantum van der Pol oscillator under the influence of external drive.}\label{energy}
\end{figure} 

The interaction term (in the absence of the drive) involves the exchange of linear excitation between the oscillators, that is the oscillator exchange energy between the $|nm\rangle$ and $|n+1,m-1\rangle$ states, where $n$ and $m$ represents the phonon number of first and second oscillator respectively. The dependence of the coupling on the occupation numbers $n$ and $m$ gives a coupling amplitude $V\sqrt{(n+1)(m+1)}$ and the corresponding energy level diagram is displayed in Fig. \ref{energy}. In our analysis we consider low excitation subspaces in the limit, $\gamma_2\gg\gamma_1$ with $N=0,1,2$ and basis states spanned by $\{|00\rangle\}$, $\{|10\rangle, |01\rangle\}$ and $\{|20\rangle$, $|11\rangle$, $|02\rangle\}$ respectively. The eigenstates of the Hamiltonian $H$ with $E=0$ are, $|0\rangle=|00\rangle$ in the ground state, $|\Psi_1^{\pm}\rangle=\frac{1}{\sqrt{2}}\left(|10\rangle\pm|01\rangle\right)$ in the first excited state. The un-normalized states $|\Psi_2^{\pm}\rangle=|20\rangle\pm|11\rangle+|02\rangle$ and $|\Psi_2^{0}\rangle=|02\rangle-|20\rangle$ are the eigenstates in the second phonon subspace.

To quantify synchronization we utilize the phase distribution and a synchronization correlator which simplifies the phase distribution into a single metric. The phase distribution of a single oscillator is 
\begin{equation}
 P(\phi_i)=\frac{1}{2\pi}\langle \phi_i|\rho|\phi_i\rangle, \label{single_phase}
\end{equation}
where $|\phi_i\rangle=\sum_{n=0}^{\infty}e^{in\phi_i}|n\rangle$ is the phase state \cite{hush, barak}. From the phase state we can construct the relative phase distribution of the oscillator, which takes the form \cite{hush, tilley}
\begin{align}
P(\varphi)&=\frac{1}{2\pi}\sum_{n,m=0}^{\infty}\sum_{k=\max(n,m)}^{\infty}e^{i\varphi(m-n)}\langle n, k-n|\rho|m, k-m\rangle, \nonumber\\
&=\frac{1}{2\pi}+\frac{1}{\pi}\text{Re}\left[\sum_{p=0}^{\infty}e^{ip\varphi}\sum_{n=0,m=0}^{\infty}\langle n, m+p|\rho|n+p,m\rangle\right], \label{phase_relative}
\end{align}
where $\varphi=\phi_2-\phi_1$ is the relative phase between the oscillators. Here we define synchronization measures as \cite{weiss}
\begin{eqnarray}
S_i&=&|S_i|e^{-\phi_i}=\frac{\langle a_i^{\dagger}\rangle}{\sqrt{\langle a_i^{\dagger}a_i\rangle}},\qquad i=1,2 \nonumber\\
S_3&=&|S_3|e^{-i\varphi}=\frac{\langle a_1^{\dagger}a_2\rangle}{\sqrt{\langle a_1^{\dagger}a_1\rangle\langle a_2^{\dagger}a_2\rangle}}, \label{synch_meas}
\end{eqnarray} 
where $S_{1,2}$ denotes the synchronization measure of the first and second oscillator with the drive respectively and $S_3$ represents the synchronization measure between the oscillators (in this work we used the steady-state solver of QuTip to numerically obtain the phase-distribution and synchronization measures \cite{johan1,johan2}).
\begin{figure}
	\centering
	\includegraphics[width=0.5\textwidth]{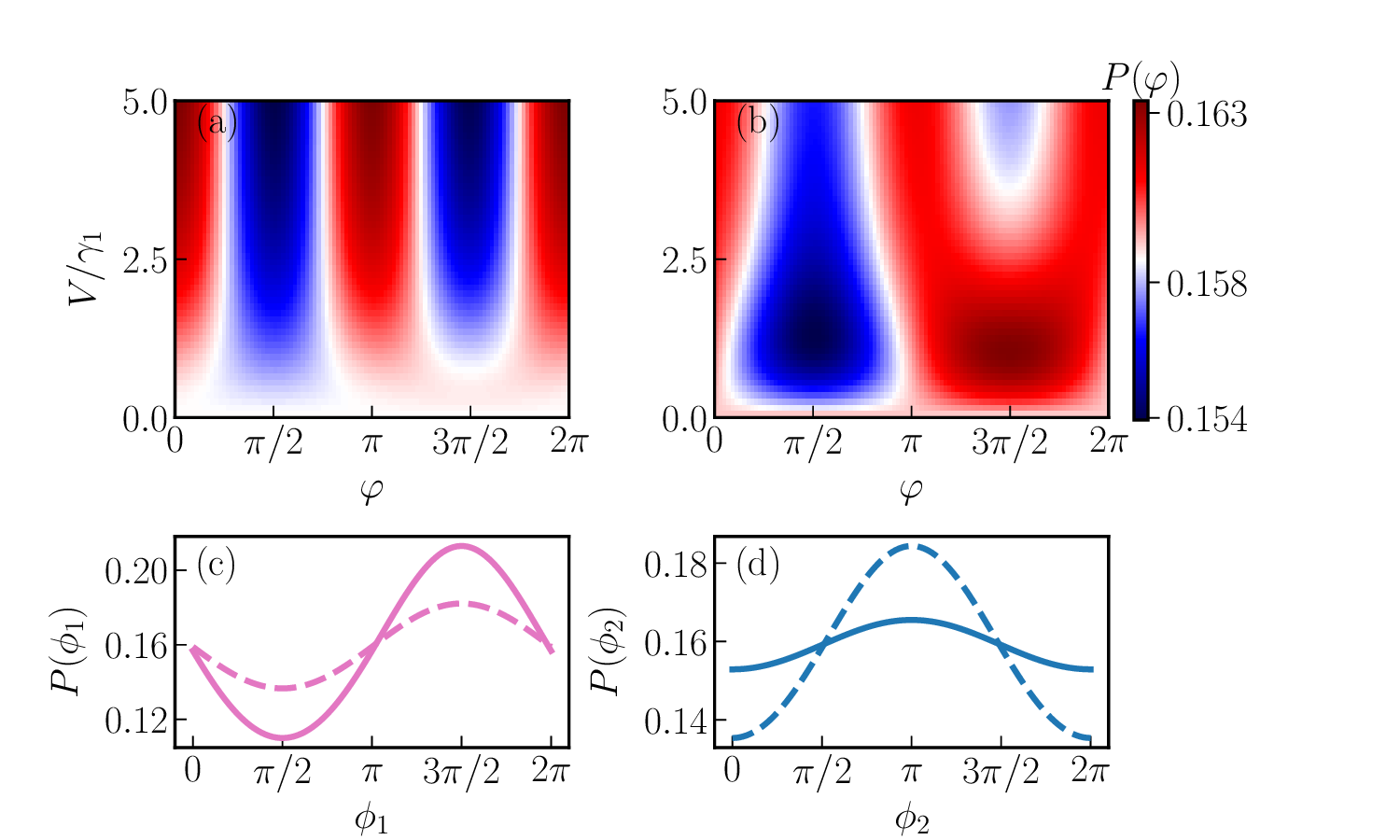}
	\caption{Relative phase distribution $P(\varphi)$ in Eq. (\ref{phase_relative}), as a function  of coupling strength $V/\gamma_1$ for (a) $E/\gamma_1=0.1$ and (b) $E/\gamma_1=0.5$ and $\Delta_1=\Delta_2=0$. Phase distribution of individual oscillators $P(\phi_1)$ and $P(\phi_2)$, see Eq. (\ref{single_phase}),  are plotted in (c)  and (d) respectively, for $V/\gamma_1=1.0$ (solid curve) and $V/\gamma_1=3.0$ (dashed curve) with $E/\gamma_1=0.5$. In all these cases we have considered $\gamma_2/\gamma_1=10$.}\label{phase_dist}
\end{figure}

 Figures \ref{phase_dist}(a) and \ref{phase_dist}(b) display the relative phase distribution $P(\varphi)$ evolved as a function of the coupling strength when both the oscillators are in resonance with the drive ($\Delta_1=0$ and $\Delta_2=0$). From Fig. \ref{phase_dist}(a) we can observe maximas at $\varphi=0$ and $\varphi=\pi$ corresponding to in-phase and anti-phase synchronization when we consider the low driving strength. A clearer picture of the phase distribution can be obtained by increasing the driving strength while maintaining the system within the low-excitation subspace ($E\le\gamma_1$), as shown in Fig. \ref{phase_dist}(b). When the coupling strength is less than the drive ($V<E$) the relative phase distribution shows a single maxima at $\varphi=3\pi/2$. When the coupling strength is increased ($V>E$) we can observe the bistable behaviour in the phase-distribution as shown in the Figs. \ref{phase_dist}(a) and \ref{phase_dist}(b) (the mathematical details are provided in the supplementary material \cite{supp}). Further insight into this behaviour can be gained by analyzing the phase distributions of the first and second oscillators, $P(\phi_1)$ and $P(\phi_2)$ respectively, in the regimes $V<E$ and $V>E$, which is illustrated in Figs. \ref{phase_dist}(c) and \ref{phase_dist}(d). From Fig. \ref{phase_dist}(c) we can observe a maxima, peaked at $\phi_1=3\pi/2$, in the phase distribution of the first oscillator $P(\phi_1)$ , while $P(\phi_2)$ shows a rather broad peak at $\phi_2=\pi$ (Fig. \ref{phase_dist}(d)) when $V<E$ (solid curve).  This implies that, in this regime, although the phase-locking of second oscillator with the drive is weak, the coupling induced phase-locking is still present, therefore the relative phase distribution of the system $P(\varphi)$ shows a maxima at $\varphi=3\pi/2$ in Fig. \ref{phase_dist}(b), which is influenced by the strong phase-locking of the first oscillator with the drive. When the coupling strength is increased ($V>E$) the phase distribution $P(\phi_1)$ shows broader peak at $\phi_1=3\pi/2$ and $P(\phi_2)$ shows a narrow peaked distribution at $\phi_2=\pi$, indicated by the dashed curve in Figs. \ref{phase_dist}(c) and \ref{phase_dist}(d). The amplitude of the phase distribution indicates the energy exchange between the oscillators at the resonance. At low coupling regime the first oscillator is strongly synchronized to the drive but as the coupling strength is increased coupling allows energy transfer to the second oscillator such that drive influences the second oscillator indirectly. This enhances the synchronization of the second oscillator with the drive. Thus, the energy transfer between the oscillators at the resonance collectively effects the synchronization of this coupled system (\ref{mas_coup}) when the coupling strength is increased.  As a result we can observe a transition from single maxima to bistable behaviour in the phase distribution of the coupled system (\ref{mas_coup}).

To analyze energy redistribution between the oscillators, we examine the steady-state probability amplitudes of the oscillator modes and illustrate this effect using the absolute value of the synchronization measures defined in Eq. (\ref{synch_meas}). Here, we adopt the Schr\"{o}dinger equation approach to calculate the probability amplitudes. Under weak driving conditions, we assume that the system (\ref{mas_coup}) occupies the lowest energy levels as shown in Fig. \ref{energy}, and the quantum state of the system can be written as 
\begin{equation}
|\psi\rangle=c_{00}|00\rangle+c_{10}|10\rangle+c_{01}|01\rangle+c_{20}|20\rangle+c_{11}|11\rangle+c_{02}|02\rangle, \label{ansatz}
\end{equation}
where $c_{nm}$ represents the probability amplitude of the corresponding quantum state with $n$ and $m$ representing first and second oscillator modes respectively. To obtain the probability amplitudes, we solve the Schr\"{o}dinger equation $i\frac{\partial |\psi\rangle}{\partial t}=H_{eff}|\psi\rangle$, where $H_{eff}=H_{0}+H_{I}-\sum_{i=1}^{2}\left(i\frac{\gamma_1}{2}a_{i}a_{i}^{\dagger}+i\frac{\gamma_2}{2}a_{i}^{\dagger 2}a_{i}^2\right)$ is the modified non-Hermitian Hamiltonian which includes the linear gain and the nonlinear loss terms. In the first excited state, we obtain the condition
\begin{equation}
c_{01} = \frac{-V}{\Delta_2 - \frac{3}{2} i \gamma_1} c_{10}. \label{proba01}
\end{equation}
 Equation (\ref{proba01}) implies that under the condition $V \gg \gamma_1, \Delta_2$, the probability of finding phonons is suppressed in the first oscillator mode and enhanced in the second. Clearly, coupling as well as the detunings $\Delta_1$ and $\Delta_2$ affect the population transfer between the oscillators.
\begin{figure}
	\includegraphics[width=0.5\textwidth]{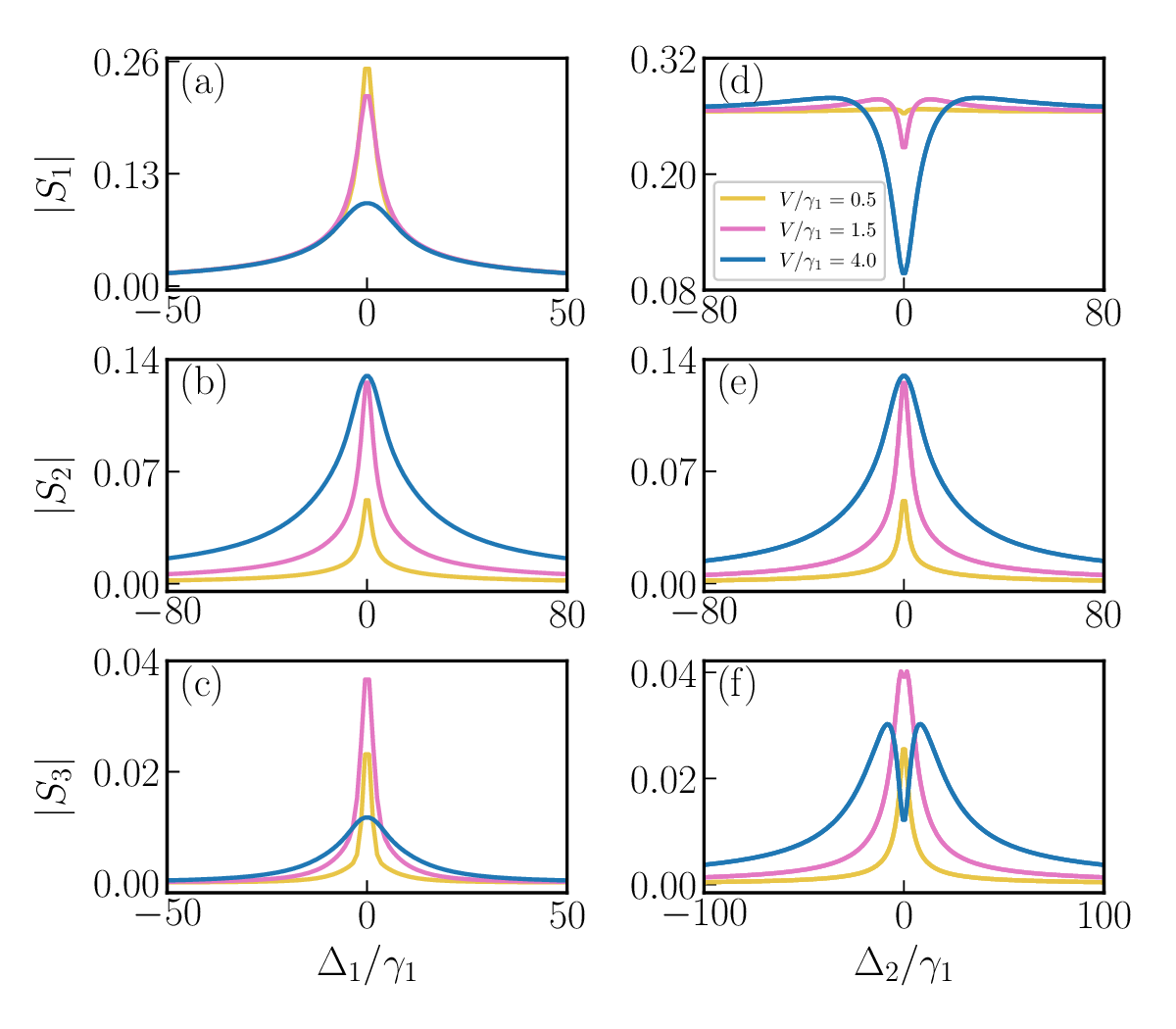}
	\caption{Absolute value of synchronization measure $S_1$ (panels (a) and (d)), $S_2$ (panels (b) and (e)) and $S_3$ (panels (c) and (f)), defined in Eq. (\ref{synch_meas}), plotted as a function of $\Delta_1$ (first column) and $\Delta_2$ (second column) for different coupling strengths (values shown in the inset of panel (d)) with $E/\gamma_1=0.5$ and $\gamma_2/\gamma_1=10$.}\label{synch2d}
\end{figure}
Figure \ref{synch2d} shows the absolute value of the synchronization measures $S_i$ as a function of the detunings $\Delta_{1,2}$ in the steady state. 

To illustrate effective population transfer, we present the synchronization behaviour of the system for two cases in Fig. \ref{synch2d}: (i) when the second oscillator is in resonance with the drive ($\Delta_2 = 0$, first column) and (ii) when the first oscillator is resonant with the drive ($\Delta_1 = 0$, second column). For $V < \gamma_1$, the condition (\ref{proba01}) implies that the phonon population in the first oscillator mode is more prominent than in the second, and the synchronization of the first oscillator with the drive is primarily governed by the direct transition with the drive. This regime is mainly characterized by dominant direct driving and the excitation through the coupling is minimal. Therefore, a strong phase-locking behaviour is observed at the resonance, $\Delta_1 = 0$, as shown in Fig.~\ref{synch2d}(a). Also, from Fig. \ref{synch2d}(d), we can observe that $|S_1|$ remains almost  constant with respect to $\Delta_2$ in the regime $V < \gamma_1$, reflecting that, when the coupling is weak, the coupling induced excitation of the first oscillator is small and its synchronization is only weakly affected by the second oscillator. The second oscillator, on the other hand, shows weak resonant peaks in response to both $\Delta_1$ and $\Delta_2$, respectively displayed in Figs. \ref{synch2d}(b) and \ref{synch2d}(e). Increasing the coupling strength to $V > \gamma_1$, amplifies the population gain in the second mode, which affects the phase-locking of both oscillators to the drive. Thus, as we can see in Fig. \ref{synch2d}(a), in response to its detuning with the drive ($\Delta_1$), the resonant peaks of the synchronization measure of the first oscillator $|S_1|$ decrease. Interestingly, we observe synchronization dips at the resonance when $|S_1|$ is varied with respect to $\Delta_2$, which become more pronounced with increasing coupling strength, as illustrated in Fig. \ref{synch2d}(d). This can be attributed to the fact that, since the coupling amplifies the phonon transfer from mode one to two, the transition pathways $|10\rangle \to |20\rangle$ (direct excitation) and $|10\rangle \to |01\rangle \to |11\rangle \to |20\rangle$ (coupling-induced excitation), as shown in Fig.~\ref{energy}, start to undergo destructive interference, which becomes more pronounced with increased coupling strength and hence the dips. The population gain in the second oscillator mode therefore enhances synchronization of the second oscillator with the drive at the resonance, as illustrated in Figs.~\ref{synch2d}(b) and~\ref{synch2d}(e), due to the coupling-induced transition to the second phonon subspace ($|10\rangle \to |01\rangle \to |11\rangle \to |02\rangle$). Thus, the destructive and constructive interference which suppresses and enhances synchronization of first and second oscillator modes to the drive is controlled by the detuning $\Delta_2$. Notably, maximal phonon transfer occurs at resonance ($\Delta_1 = \Delta_2 = 0$), while energy exchange is suppressed for $|\Delta_2| > V$ (Eq. \ref{proba01}), explaining the constant synchronized response of $|S_1|$ with respect to $\Delta_2$ in this regime (Fig.~\ref{synch2d}(d)). This interference mechanism between the transition pathways enables spectral control over mutual synchronization. The response of absolute value of mutual synchronization measure $S_3$ in response to $\Delta_1$ and $\Delta_2$ are respectively presented in Figs. \ref{synch2d}(c) and \ref{synch2d}(f). We can see resonant peaks in Fig. \ref{synch2d}(c), which increases and then decreases with coupling strength. In response to $\Delta_1$ and $\Delta_2$, the coupled system shows synchronized peaks in the regime $V<\gamma_1$. This suggests that, eventhough the driving effect is dominant in this regime, the finite coupling between the oscillators still facilitate the transfer of phase coherences and energy between the oscillators. Consequently, the second oscillator weakly synchronize with the first as evidenced by the resonant peaks in Figs. \ref{synch2d}(c) and \ref{synch2d}(f) (yellow curves). This indicates that the synchronization still persists in the weak coupling regime. When coupling strength is increased the constructive and destructive mechanism splits the energy spectrum into two normal modes as shown in Fig. \ref{energy}. Hence the bare eigenstates $|10\rangle$ and $|01\rangle$ are converted into collective states $|\Psi_1^{\pm}\rangle$ with eigenenergies $e_1^{\pm}=\pm V$ and, the transition from $|0\rangle$ to $|\Psi_1^{\pm}\rangle$ induces spectrally resolved synchronization peaks in the off-resonant regime and a dip at $\Delta_2=0$ as illustrated in Fig. \ref{synch2d}(f). The dip arises due to the destructive interference between the transition paths $|0\rangle$ to $|\Psi_1^{-}\rangle$ and $|0\rangle$ to $|\Psi_1^{+}\rangle$ inducing a normal mode induced blockade in the synchronization of the coupled oscillator.
\begin{figure}
	\centering
	\includegraphics[width=0.5\textwidth]{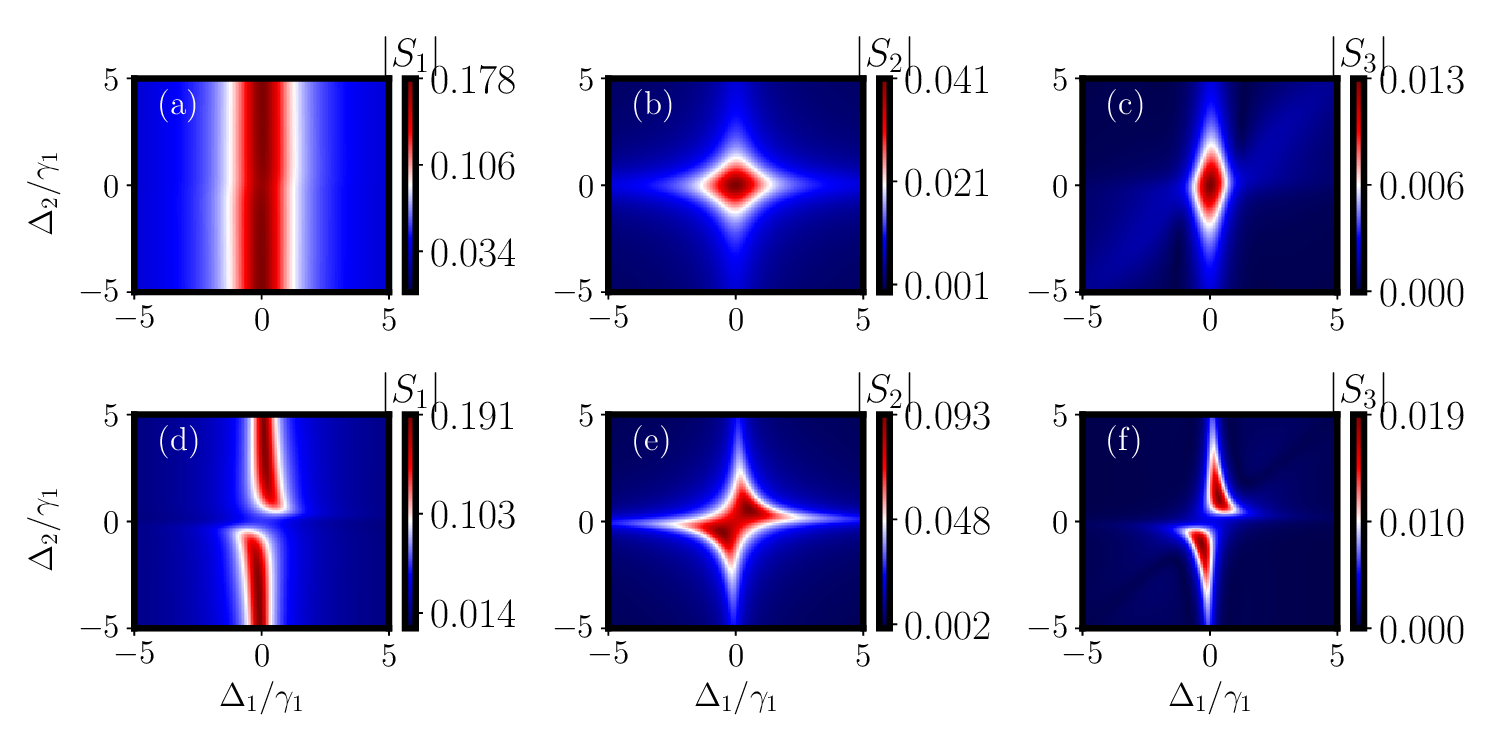}
	\caption{Synchronization measures $|S_i|$, $i=1,2,3$ (Eq. (\ref{synch_meas})), as a function of $\Delta_1$ and $\Delta_2$. Panels (a)-(c) are plotted for coupling strength $V/\gamma_1=0.6$ and panels (d)-(f) are plotted for coupling strength $V/\gamma_1=4.0$ with $E/\gamma_1=0.5$ and $\gamma_2/\gamma_1=10$.}\label{synch}
\end{figure}

Figure \ref{synch} displays the impact of both detunings, $\Delta_1$ and $\Delta_2$, on the absolute value of the synchronization measures $S_i$. In Figs. \ref{synch}(a) - \ref{synch}(c), we plot $|S_i|$ in the regime where phonon occupation dominates in oscillator mode one ($V < \gamma_1$). A vertical band structure is observed for $|S_1|$ around $\Delta_1 = 0$ as seen in Fig. \ref{synch}(a), indicating strong phase-locking of the first oscillator to the drive, with minimal variation as $\Delta_2$ is changed. In this regime, we also observe a bright synchronization region centered near resonance ($\Delta_1 = 0, \Delta_2 = 0$) for $|S_2|$ and $|S_3|$ in Figs. \ref{synch}(b) and \ref{synch}(c), respectively. Figures \ref{synch}(d) - \ref{synch}(f) illustrate the synchronization behaviour when energy exchange is amplified by coupling ($V > \gamma_1$). In Fig. \ref{synch}(d), a split in the synchronization regime of the first oscillator appears across $\Delta_2 = 0$, induced by destructive interference. This splitting becomes more pronounced as the coupling strength increases. In the off resonant regime ($|\Delta_2| > V$), the first oscillator remains strongly synchronized with the drive. Conversely, enhanced energy transfer leads to increased synchronization of the second oscillator with the drive around resonance in Fig. \ref{synch}(e), due to constructive interference enabling transitions to the second phonon subspace. Far from resonance, the synchronization region narrows. For the mutually synchronized case, Fig. \ref{synch}(f) reveals mode splitting in the synchronization regime near resonance, arising from transitions to spectrally resolved modes due to interference mechanism between oscillator modes. Away from resonance, the synchronization regime is influenced by both detunings, and the intensity of the resonant peaks decreases as one moves further away from resonance, as shown in Fig.~\ref{synch}(f).
\begin{figure}
	\centering
	\includegraphics[width=0.5\textwidth]{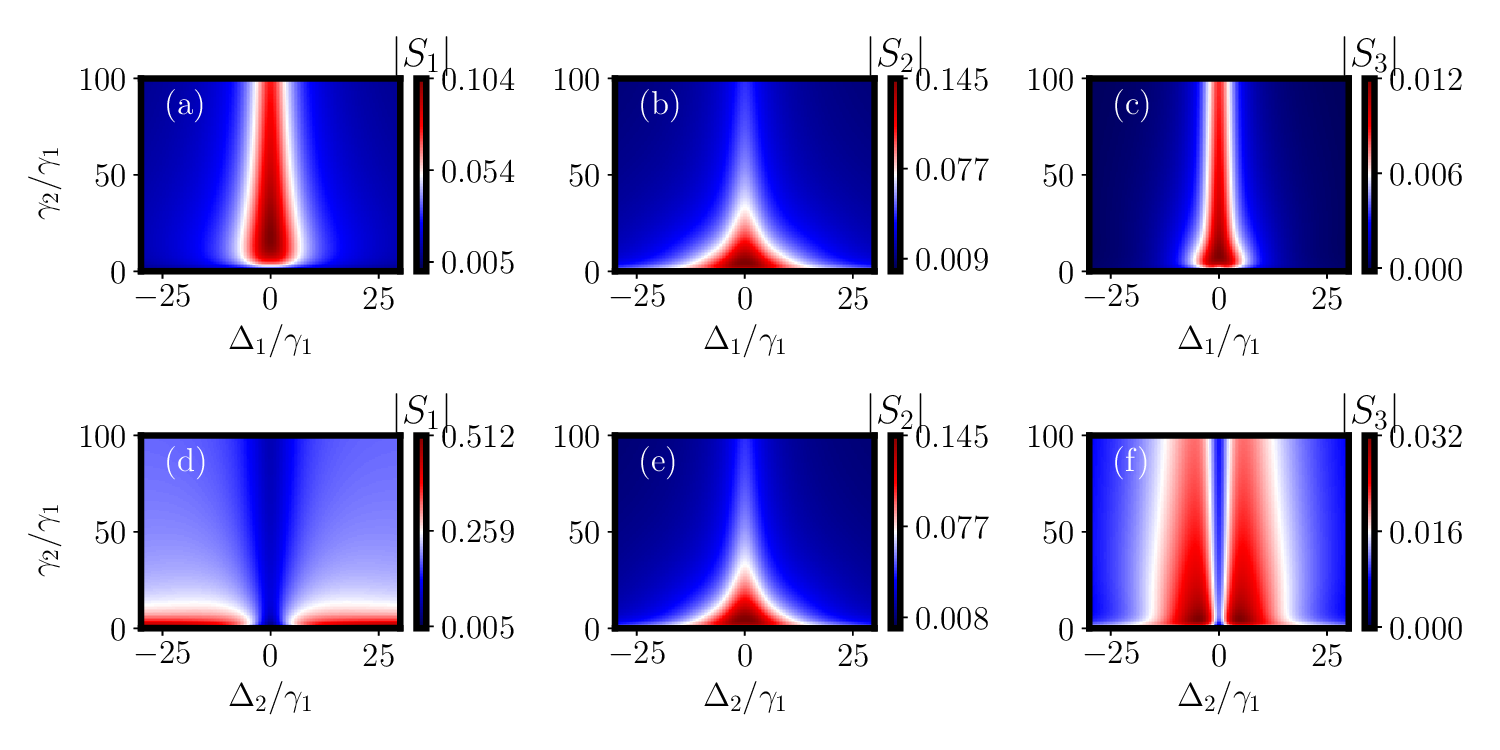}
	\caption{In the first row panels (a)-(c) synchronization measures $|S_i|$, $i=1,2,3$, are plotted as a function of $\Delta_1$ and $\gamma_2/\gamma_1$ with $\Delta_2=0$. In the second row panels (d)-(f) synchronization measures $|S_i|$ ($i=1,2,3$), defined in Eq. (\ref{synch_meas}), are plotted as a function of $\Delta_2$ and $\gamma_2/\gamma_1$ with $\Delta_1=0$. In all these figures we consider $V/\gamma_1=5.0$ and $E/\gamma_1=0.5$.}\label{damp}
\end{figure}

In our analysis, we considered $\gamma_2 \gg \gamma_1$ to ensure population remains confined to the lowest Fock states ($N \le 2$). Figure \ref{damp} demonstrates the impact of $\gamma_2/\gamma_1$ on the synchronization measures $S_i$. As the rate $\gamma_2/\gamma_1$ increases the system predominantly occupies lower Fock states ($N < 2$), and we observe that the synchronization of the first oscillator with the drive ($|S_1|$) and mutual synchronization between the oscillators ($|S_3|$) persist (Figs. \ref{damp}(a) and \ref{damp}(c)) \cite{supp}. However, the second oscillator's synchronization with the drive diminishes at high $\gamma_2/\gamma_1$ (Figs. \ref{damp}(b) and \ref{damp}(e)), indicating that its synchronization requires population in higher Fock states ($N \geq 2$), which are suppressed by strong dissipation rates $\gamma_2$. Notably, the synchronization blockade (Figs.~\ref{damp}(d) and \ref{damp}(f)) persists even for large $\gamma_2/\gamma_1$, as it arises from transitions between the ground and first excited Fock states. While the blockade mechanism shows slight broadening with increasing $\gamma_2/\gamma_1$, the interference-induced normal mode splitting in the synchronization regime remains robust even for $\gamma_2 \gg V$, confirming its resilience against strong dissipation. Our results indicate that, the mechanism underlying the synchronization blockade and the associated normal-mode splitting in the low occupation manifold remains robust against strong dissipation, whereas the synchronization features of the second oscillator ($|S_2|$), which requires occupation of higher Fock state, are more sensitive to increasing $\gamma_2$. This distinction explains why the core blockade effect is resilient against the increasing dissipation rates, even though the synchronization measure, $|S_2|$, does not exhibit the same robustness at large $\gamma_2/\gamma_1$.

\textit{Experimental realization:} System (\ref{mas_coup}) can be implemented in a trapped ion setup using the sideband transition of two motional modes. The dynamics of system (\ref{mas_coup}) can be achieved by facilitating the blue-sideband transition of each mode with appropriate detunings to control the linear phonon gain ($\gamma_1$) and nonlinear loss ($\gamma_2$) \cite{supp}. By additionally driving sideband transitions involving both modes with suitable detuning the effective interaction ($H_I=V(a_1^{\dagger}a_2+a_1a_2^{\dagger})$) can be generated, while an RF drive acts on the first oscillator \cite{lee}. Alternatively, the quantum van der Pol oscillator can be realized in optomechanical ``membrane-in-the-middle" setup as well. Quadratic optomechanical coupling enables two phonon absorption and emission, a laser drive on the red two-phonon sideband realizes nonlinear damping \cite{supp}, while a linearly couled cavity mode driven on the blue one-phonon sideband provides effecive negative damping \cite{walter14}. To implement the system (\ref{mas_coup}), two mechanical van der Pol oscillators interact via phonon-hopping interaction ($J$) and are coupled to a common cavity through radiation pressure ($g$) \cite{mari, xu, kuzyk}. When one of the mechanical modes is driven ($\Omega$), the Hamiltonian of the system in the frame rotating at the laser frequency can be represented in the form, 
$H=\omega_c a^{\dagger}a+\sum_{i=1,2}\Delta_{m,i}b_i^{\dagger}b_i+ga^{\dagger}a(b_i^{\dagger}+b_i)+J(b_1^{\dagger}b_2+b_1b_2^{\dagger})+\Omega(b_1^{\dagger}+b_1)$,
where $\Delta_{m,1}=\omega_{m,1}-\omega_d$ and $\Delta_{m,2}=\omega_{m,2}-\omega_d$ are the detunings of the mechanical mode frequencies with respect to the driving frequency.

To conclude, in this work, we have explored the synchronization dynamics of two reactively coupled quantum van der Pol oscillators, revealing the emergence of a normal mode splitting-induced synchronization blockade. By analysing the relative phase distribution between the oscillators we have shown that the system undergoes a transition from a single peaked distribution in low coupling regime to bistable distribution in high-coupling regime revealing blockade mechanism in the considered coupled system. Also, by examining the energy spectrum and from the obtained analytical results of the eigen-system, we have demonstrated that the coupled system undergo normal-mode splitting induced synchronization blockade through destructive and constructive interference mechanism in the oscillator modes.  We observed that system's dependence on its parameters offers insights into controlling and manipulating synchronization in the quantum van der Pol oscillator. Also, we found that strong dissipation stabilizes the blockade by confining the population to lower Fock states. Our results can contribute to the growing understanding of quantum synchronization and its potential applications in quantum technologies, such as quantum information processing and quantum sensing.

NT and MS acknowledges the support from Science and Engineering Research Board,  Government of India, under the Grant No. CRG/2021/002428.

\end{document}


	\title{Supplemental Material}
	\maketitle
\section{Perturbation Analysis and Phase distribution}
\begin{figure}
	\centering
	\includegraphics[width=0.7\textwidth]{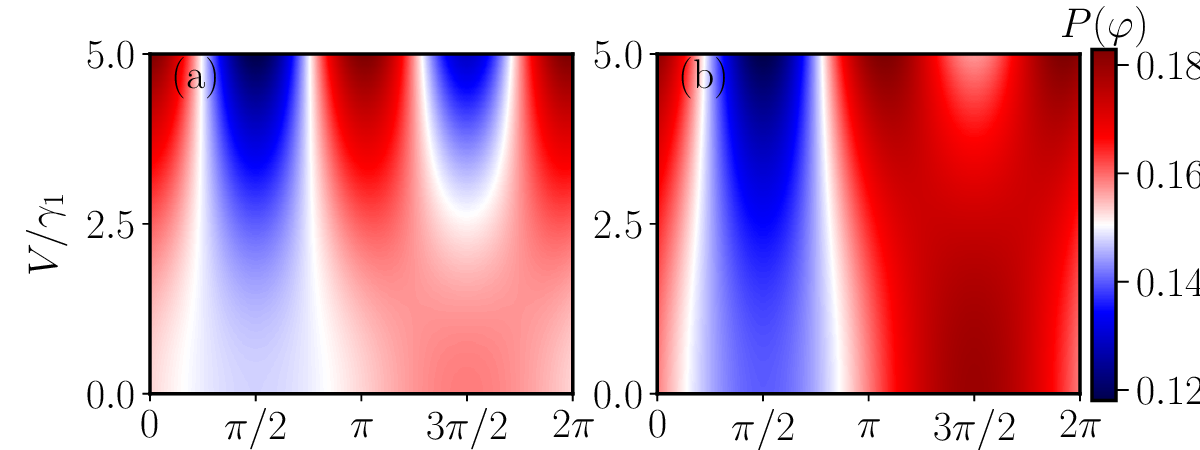}
	\caption{Relative phase distribution $P(\varphi)$ obtained in Eqs. (S.5) and (S.7), as a function  of coupling strength $V/\gamma_1$ for (a) $E/\gamma_1=0.1$ and (b) $E/\gamma_1=0.5$ and $\Delta_1=\Delta_2=0$. Here we have considered $\gamma_2/\gamma_1=10$.}\label{Fig:S1}
\end{figure}

In this supplemental material, we present the analytical expression of the Phase distribution in Eq. (4), using perturbation analysis. Perturbation analysis is a suitable approach to calculate the steady-state density matrix of the system (1), through which we can have insight on how the phase distribution of the system acts. To obtain the steady-state density matrix we apply the number state basis, $\rho_{nm}^{(p)}=\langle n, m+p|\rho|n+p,m\rangle$ in the limit, $V, E \ll \gamma_1+\gamma_2$, to the master equation (1) with the interaction term (2) which leads to
\begin{equation}
\dot{\rho}_{nm}^{(p)}=\lambda_{n,m}^{(p)}\rho_{n,m}^{(p)}-(\nu_1^{(p)}+\nu_2^{(p)})\rho_{n,m}^{(p)}, \label{stea_mas} \tag{S.1}
\end{equation}
where
\begin{align}
\nu_1^{(p)}&=iE\left(\sqrt{n+1}\rho_{n+1,m}^{(p-1)}-\sqrt{n+p}\rho_{n,m}^{(p-1)}\right),\nonumber\\
\nu_2^{(p)}&=iV\left(\sqrt{(n+1)(m+p)}\rho_{n+1,m}^{(p-1)}-\sqrt{(n+p)(m+1)}\rho_{n,m+1}^{(p-1)}\right),\nonumber\\
\lambda_{n,m}^{(p)}&=i(\Delta_1-\Delta_2)p-\Gamma_{n,m}^{(p)}, \tag{S.2}  \label{steady_state} 
\end{align}
with $\Gamma_{n,m}^{(p)}=\gamma_1\left(n+m+p+2\right)+\gamma_2\left(n(n-1)+m(m-1)+(n+p)(n+p-1)+(m+p)(m+p-1)\right)$ and the equations for $\nu_{1,2}$ arises from the interaction term. The $p^{th}$ order correction to steady-state is then given by the expression
\begin{equation}
\rho_{nm}^{(p)}=\frac{1}{\lambda_{n,m}^{(p)}}\left(\nu_1^{(p)}+\nu_2^{(p)}\right). \tag{S.3} \label{pthcorr}
\end{equation}
In very weak coupling regime ($V<\gamma_1$) we have the lowest order contribution of the interaction terms to the density matrix such that
\begin{align}
\nu_1^{(1)}&=iE\sqrt{n+1}\left(\rho_{n+1,m}^{(0)}-\rho_{n,m}^{(0)}\right),\nonumber\\
\nu_2^{(1)}&=iV\sqrt{(n+1)(m+1)}\left(\rho_{n+1,m}^{(0)}-\rho_{n,m+1}^{(0)}\right). \tag{S.4}\label{inter1}
\end{align}
From the above expression we obtain the first order correction to density matrix $\rho_{n,m}^{(1)}$ and its contribution to phase distribution (Eq. (4)) is 
\begin{equation}
P(\varphi)=\frac{1}{2\pi}+\frac{1}{\pi}(c_0V+c_1E)\sin\varphi. \tag{S.5}
\end{equation}
We note that, when $E=0$ the first order components of the steady-state density matrix satisfies the relation $\rho_{n,m}^{(1)}=-\rho_{m,n}^{(1)}$. In the coupling regime $V>\gamma_1$, since the coupling term dominates the drive term, the first order components of the density matrix approximately satisfies the relation $\rho_{n,m}^{(1)}\approx-\rho_{m,n}^{(1)}$ such that its contribution to phase distribution (4) becomes negligible. Therefore, the second order components ($p=2$) of the interaction terms are given by
\begin{align}
\nu_1^{(2)}&=iE\left(\sqrt{n+1}\rho_{n+1,m}^{(1)}-\sqrt{n+2}\rho_{n,m}^{(1)}\right),\nonumber\\
\nu_2^{(2)}&=iV\left(\sqrt{(n+1)(m+2)}\rho_{n+1,m}^{(1)}-\sqrt{(n+2)(m+1)}\rho_{n,m+1}^{(1)}\right), \tag{S.6} \label{inter2}
\end{align}
contributes to the phase distribution in Eq. (4) resulting in 
\begin{equation}
P(\varphi)=\frac{1}{2\pi}+\frac{1}{\pi}(c_2V^2+c_3EV+c_4E^2)\cos 2\varphi, \tag{S.7}\label{strong_phase}
\end{equation}
where $c_i$'s are the constants depending on the damping parameters. In Fig. \ref{Fig:S1} we have displayed the phase distribution analytically obtained in Eqs. (S.5) and (S.7). As stated above, for a finite value of drive strength, phase distribution shows a single maxima at $\varphi=3\pi/2$ in the weak coupling regime and it becomes bistable in the regime where $V>\gamma_1$. Therefore, we can see that both the analytical and numerical results given in the main text qualitatively agree with each other.

\section{Synchronization features at low dissipation rates and experimental realizations of enhanced nonlinear damping}
\begin{figure}
	\centering
	\includegraphics[width=0.7\textwidth]{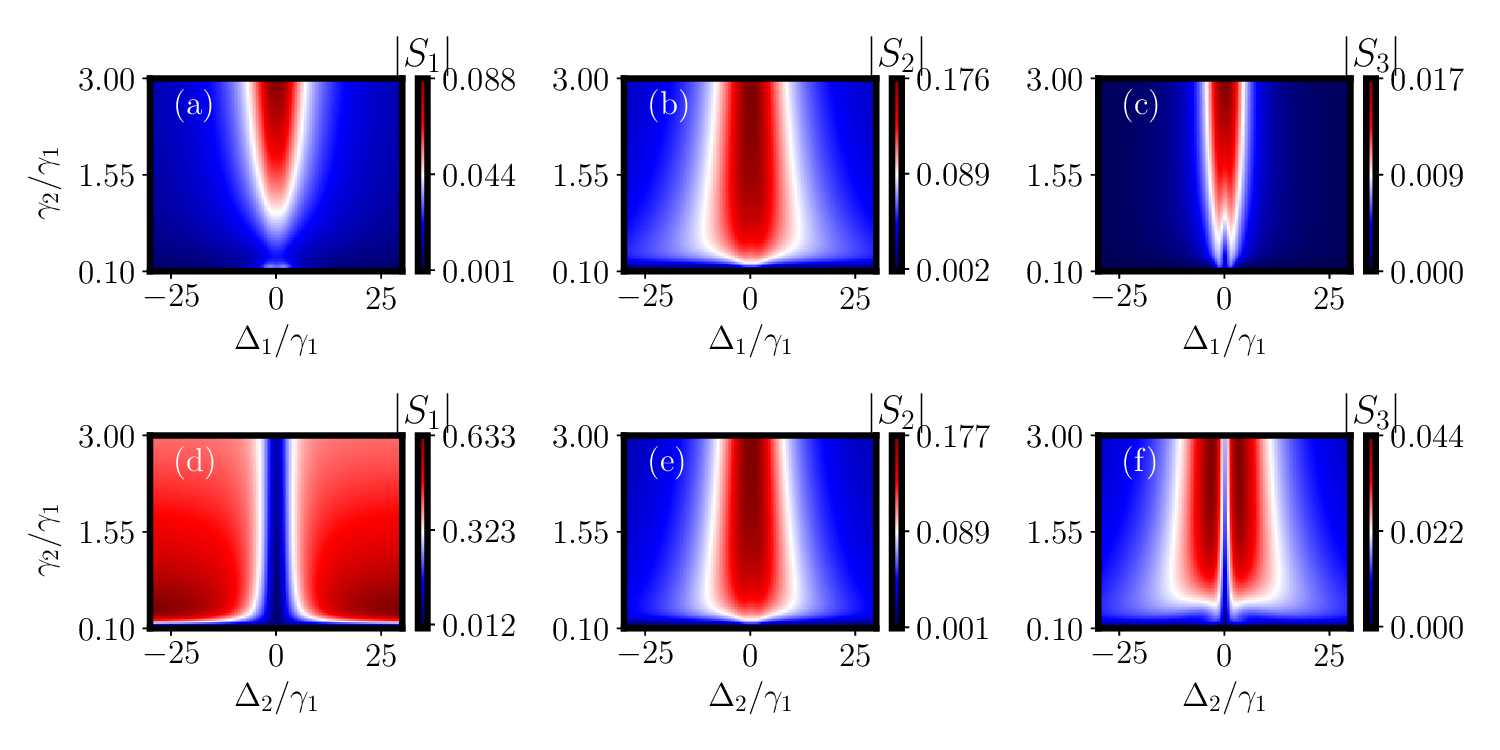}
	\caption{In the first row panels (a)-(c) synchronization measures $|S_i|$, $i=1,2,3$, are plotted as a function of $\Delta_1$ and $\gamma_2/\gamma_1$ with $\Delta_2=0$. In the second row panels (d)-(f) synchronization measures $|S_i|$ ($i=1,2,3$), defined in Eq. (5), are plotted as a function of $\Delta_2$ and $\gamma_2/\gamma_1$ with $\Delta_1=0$. In all these figures we consider $V/\gamma_1=3.0$ and $E/\gamma_1=0.5$.}\label{damp_revised}
\end{figure}

In many experimental platforms the nonlinear damping rate is often weaker than linear dissipation rate $\gamma_1$. Figure \ref{damp_revised} demonstrates the response of synchronization measures $|S_{1,2,3}|$ as function of $\gamma_2/\gamma_1$ and detunings $\Delta_{1,2}$ for lower range of dissipation rates. For lower values of $\gamma_2/\gamma_1$ the system is no longer restricted to lowest Fock state and higher excited state acquires appreciable population. We can observe that in the low dissipation regime, where the coupling dominates over the damping, the coupling induced synchronization of second oscillator with the drive is enhanced as illustrated in Figs. \ref{damp_revised}(b) and \ref{damp_revised}(e) while the direct synchronization of first oscillator with the drive becomes less pronounced as shown in Figs. \ref{damp_revised}(a) and \ref{damp_revised}(d) since the coupling facilitates the population transfer to the second oscillator mode. For the collective behaviour, the interference induced splitting still persists as shown in Figs. \ref{damp_revised}(c) and \ref{damp_revised}(f) indicating that the blockade remains visible for lower values of $\gamma_2/\gamma_1$.

Although the value of nonlinear damping $\gamma_2$ is naturally smaller than linear dissipation rate $\gamma_1$,  several recent experiments demonstrated that, large $\gamma_2$ is achievable. In the trapped ion systems, quadratic dissipation of the form $\mathcal{D}[a^2]$ has been realized using reservoir engineering protocols based on sideband coupling and optical pumping, enabling experimental implementation and synchronization of quantum van der Pol oscillator \cite{li}. Related schemes in circuit QED architectures employs two-phonon loss to stabilize bosonic codes  and control bit-flips times, effectively realizing a large nonlinear damping rate \cite{marquet1, berdou, touzard, marquet2}. In optomechanical platforms, quadratic optomechanical interactions in the reversed-dissipation regime provide another route to engineer nonlinear mechanical damping \cite{lee}. These works support the feasibility of accessing regimes with $\gamma_2\geq\gamma_1$ assumed in our analysis.
